\documentstyle[twoside,fleqn,epsfig,espcrc2]{article}

\newcommand{\AmS}{{\protect\the\textfont2
  A\kern-.1667em\lower.5ex\hbox{M}\kern-.125emS}}

\def\ap#1#2#3   {{\em Ann. Phys. (NY)} {\bf#1} (#2) #3.}
\def\apj#1#2#3  {{\em Astrophys. J.} {\bf#1} (#2) #3.}
\def\apjl#1#2#3 {{\em Astrophys. J. Lett.} {\bf#1} (#2) #3.}
\def\app#1#2#3  {{\em Acta. Phys. Pol.} {\bf#1} (#2) #3.}
\def\ar#1#2#3   {{\em Ann. Rev. Nucl. Part. Sci.} {\bf#1} (#2) #3.}
\def\cpc#1#2#3  {{\em Computer Phys. Comm.} {\bf#1} (#2) #3.}
\def\err#1#2#3  {{\it Erratum} {\bf#1} (#2) #3.}
\def\ib#1#2#3   {{\it ibid.} {\bf#1} (#2) #3.}
\def\jmp#1#2#3  {{\em J. Math. Phys.} {\bf#1} (#2) #3.}
\def\ijmp#1#2#3 {{\em Int. J. Mod. Phys.} {\bf#1} (#2) #3.}
\def\jetp#1#2#3 {{\em JETP Lett.} {\bf#1} (#2) #3.}
\def\jpg#1#2#3  {{\em J. Phys. G.} {\bf#1} (#2) #3.}
\def\mpl#1#2#3  {{\em Mod. Phys. Lett.} {\bf#1} (#2) #3.}
\def\nat#1#2#3  {{\em Nature (London)} {\bf#1} (#2) #3.}
\def\nc#1#2#3   {{\em Nuovo Cim.} {\bf#1} (#2) #3.}
\def\nim#1#2#3  {{\em Nucl. Instr. Meth.} {\bf#1} (#2) #3.}
\def\np#1#2#3   {{\em Nucl. Phys.} {\bf#1} (#2) #3.}
\def\pcps#1#2#3 {{\em Proc. Cam. Phil. Soc.} {\bf#1} (#2) #3.}
\def\pl#1#2#3   {{\em Phys. Lett.} {\bf#1} (#2) #3.}
\def\prep#1#2#3 {{\em Phys. Rep.} {\bf#1} (#2) #3.}
\def\prev#1#2#3 {{\em Phys. Rev.} {\bf#1} (#2) #3.}
\def\prl#1#2#3  {{\em Phys. Rev. Lett.} {\bf#1} (#2) #3.}
\def\prs#1#2#3  {{\em Proc. Roy. Soc.} {\bf#1} (#2) #3.}
\def\ptp#1#2#3  {{\em Prog. Th. Phys.} {\bf#1} (#2) #3.}
\def\ps#1#2#3   {{\em Physica Scripta} {\bf#1} (#2) #3.}
\def\rmp#1#2#3  {{\em Rev. Mod. Phys.} {\bf#1} (#2) #3.}
\def\rpp#1#2#3  {{\em Rep. Prog. Phys.} {\bf#1} (#2) #3.}
\def\sjnp#1#2#3 {{\em Sov. J. Nucl. Phys.} {\bf#1} (#2) #3.}
\def\spj#1#2#3  {{\em Sov. Phys. JEPT} {\bf#1} (#2) #3.}
\def\spu#1#2#3  {{\em Sov. Phys.-Usp.} {\bf#1} (#2) #3.}
\def\zp#1#2#3   {{\em Zeit. Phys.} {\bf#1} (#2) #3.}
\newcommand{\bc}{\begin{center}}
\newcommand{\ec}{\end{center}}
\newcommand{\bi}{\begin{itemize}}
\newcommand{\ei}{\end{itemize}}

\newcommand{\tm}{\mbox{$\tau^{-}$}}
\newcommand{\nut}{\mbox{$\nu_{\tau}$}}

\newcommand{\nue}{\mbox{$\nu_{e}\,$}}

\newcommand{\dms}{\mbox{$\Delta m^{2}$}}
\newcommand{\Pt}{\mbox{$P_T$}}
\newcommand{\Ptm}{\mbox{$P_T^m$}}

\def\mutau{\mbox{$\;\nu_{\mu} \rightarrow  \nu_{\tau}  \;$}}
\def\etau{\mbox{$\;\nu_{e} \rightarrow  \nu_{\tau}  \;$}}

\def\numu{\mbox{$\;\nu_{\mu}\;$}}
\title{The NOMAD Experiment : Status Report} 

\author{Marco Laveder\address{Dipartimento di Fisica `` G.Galilei",
        University of Padova and \\
        INFN Sezione di Padova, I-35131 Padova, Italy}%
        \thanks{Presented on behalf of the NOMAD Collaboration.}
}

\begin{document}
\newcommand{\fcaption}[1]{
        \refstepcounter{figure}
        \setbox\@tempboxa = \hbox{\tenrm Fig.~\thefigure. #1}
        \ifdim \wd\@tempboxa > 6in
           {\begin{center}
        \parbox{6in}{\tenrm\baselineskip=12pt Fig.~\thefigure. #1 }
            \end{center}}
        \else
             {\begin{center}
             {\tenrm Fig.~\thefigure. #1}
              \end{center}}
        \fi}

\newcommand{\tcaption}[1]{
        \refstepcounter{table}
        \setbox\@tempboxa = \hbox{\tenrm Table~\thetable. #1}
        \ifdim \wd\@tempboxa > 6in
           {\begin{center}
        \parbox{6in}{\tenrm\baselineskip=12pt Table~\thetable. #1 }
            \end{center}}
        \else
             {\begin{center}
             {\tenrm Table~\thetable. #1}
              \end{center}}
        \fi}

\begin{abstract}
\noindent
The NOMAD experiment has been designed to search for
$\nu_\tau$ appearance in the CERN wide-band neutrino
beam . 
The detector is now completed and has been further improved. 
All subdetectors are working well.
The experiment, where the search for oscillation
is based on kinematical criteria, 
will reach the sensitivity
$\dms~>~0.7\,$ eV$^{2}$ for maximal mixing and
$\dms~>~50\,$ eV$^{2}$ for mixing angles $\sin^2 2\theta > 3.8\times 10^{-4}$
after 2 years of running, making possible to explore
a region of cosmological interest.
Preliminary measurements are presented from the 1994 and 1995 data
samples.

\end{abstract}

\pagestyle{plain}
\setcounter{page}{1}
 
\newpage
\maketitle

\section{Introduction}
 
The goal of the NOMAD experiment \cite{nomad_prop}
are neutrino oscillations \mutau.
The experiment searches for the appearance
of tau neutrinos ,\nut{}, in the CERN SPS wide band neutrino beam,
which consists mostly of muon neutrinos \numu.
It is therefore an appearance experiment.
The search is based on the detection
of \tm production and decay:

\begin{displaymath}
\begin{array}{ccr@{}lcrcl}
\nu_{\tau} + N &\to&X+{}&{\tau^-}{}&&&&\\
&&&{\tau^-}{}&\to& {\tau^-} { decay \, \,mode}&&\\
\end{array}
\end{displaymath}

The natural contamination of \nut\ 
in the SPS neutrino beam 
is negligible ($\simeq 10^{-7}$),therefore
\tm eventually observed
could only result
from \mutau or \etau oscillations.
\par
Neutrino oscillations can take place if the states 
of the neutrinos produced in the weak interaction
processes are not stationary, but rather are
superpositions of stationary states of neutrinos
having different nonzero masses .
\par
For massive $\nu \,$, the flavour eigenstates 
$\nu_{e} \,$ ,$ \nu_{\mu} \,$, $ \nu_{\tau} \,$, 
can be expressed in terms of the mass eigenstates 
$\, \nu_1 \,$ , $ \nu_2 \, $, $\nu_3 \, $,
through an unitary mixing matrix.
\par 
In the simplified hypothesis
of a mixing between two neutrino families,
the probability of observing \mutau oscillations at
a distance $L$
is given by
\begin{equation}
P = \sin^2 2\theta \sin^2 
1.27 \frac{\Delta m^2 ({ eV}^2) L({ km})}{ E_{\nu}({ GeV})}
\end{equation}
where $E_\nu$ is the neutrino
energy in GeV ,$\theta $ the mixing angle, 
$\Delta m^2 =\mid m_1^2 - m_2^2 \mid $
the eigenstates mass squared difference in eV$^2$
and $L$ is the distance in Km between the neutrino
production and observation.

\begin{figure}[hbt]
  \bc
  \hspace{0mm}  \epsfig{file=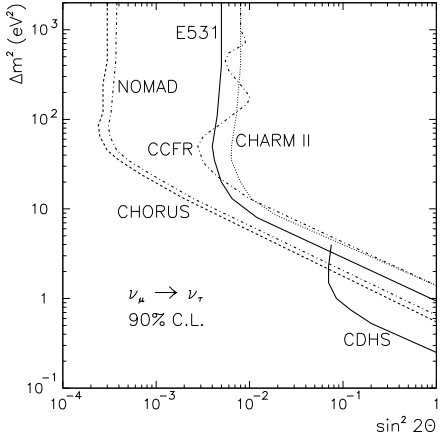,width=58mm}
  \ec
  \caption {Existing limits in the neutrino oscillation parameter 
space\protect\cite{EFIVE,CDHS,CHARMtwo,CCFR} compared to
the one achievable by on-going experiments\protect\cite{nomad_prop,CHORUS}
after two years of running.}
  \label{pspace_osc}
\end{figure}

After two years of running ,equivalent to an integral
of $2.4\times 10^{19}$ protons-on-target (pot),
a sample of about 1.5~million neutrino interactions
will be collected within NOMAD active target.
The analysis of these events will allow probing neutrino
oscillations in a wide region of the parameter space 
($\sin^2 2\theta,\Delta m^2 $),
see Fig.~\ref{pspace_osc}, exploring a mass region of 
cosmological interest :\\

\noindent
$ m_{\nu_\tau} \geq 0.8 $ { eV  for } $ \sin^2 2\theta = 1 .$\\ 
$ m_{\nu_\tau} \geq 7.0 $ { eV  for } $ 
\sin^2 2\theta \geq 3.8 \times 10^{-4}.$\\

In fact if tau neutrinos have a mass in the region
of  10\,\, eV,
they are suitable candidates for hot dark matter .

\par
While the existence of the tau neutrino in this mass
region was already investigated by previous
experiments (see for example 
Refs\cite{EFIVE,CDHS,CHARMtwo,CCFR}),
NOMAD improves the
most stringent limit given by the
E531 exp. \cite{EFIVE} by one order of magnitude.

\section{The neutrino beam}
The CERN SPS neutrino Wide Band Beam (WBB) is produced using 450 GeV
protons  hitting a Be target with a cycle
of 14.4 s. 
The intensity delivered is about $2\times 10^{13}$ protons
per cycle.
\par
The NOMAD 
detector is located downstream the target at about 820~m .
The relative abundances of the different species of neutrinos
and their average energies are listed in
Table \ref{tab:neuene}.
\begin{table}[b]
\setlength{\tabcolsep}{1.225pc}
\newlength{\digitwidth} \settowidth{\digitwidth}{ 0}
\catcode`?=\active \def?{\kern\digitwidth}
\caption{CERN WBB averaged neutrino energies}
\label{tab:neuene}
\begin{tabular}{lcc}
\hline
Neutrino & $<E_\nu>$ $({ GeV})$ & r.a.\\
\hline
$\nu_\mu$ & 26.9 & 1.0 \\
$\bar{\nu}_\mu$ & 21.7 & 0.06 \\
$\nu_e$ & 47.9 & 0.007 \\
$\bar{\nu}_e$ & 35.3 & 0.002 \\
\hline
\end{tabular}
\end{table}
\par
The sensitivity of the experiment
is calculated for an integrated intensity
of $2.4\times 10^{19}$
pot which corresponds to about:
$1.1\times 10^{6}$  $\nu_\mu$ $CC$ events,
$3.7\times 10^{5}$ $\nu_\mu$ $NC$ events and
$1.3\times 10^{4}$ $\nu_e$ $CC$ events in the fiducial
target of the NOMAD detector (mass = 2.7 tons,
area = $2.6 \times 2.6 \,$ m$^2$).

\section{Neutrino tau search}
NOMAD aims at identifying \tm
production and decay using suitable kinematical  selection criteria
such as missing $\Pt$, angular correlations etc.
To do so, very good energy, momentum and angular
resolution are needed.
The \tm is detected through a large variety of
its decay modes listed in Table \ref{tab:sumchan}.
As an example of the kinematical analysis the electronic
decay of \tm will be considered.
The background to this channel
is caused by \nue charged current events
which constitute $\approx 1\%$
of the total number of neutrino interactions.

\begin{table}[hbt]
\setlength{\tabcolsep}{1.5pc}
\settowidth{\digitwidth}{ 0}
\catcode`?=\active \def?{\kern\digitwidth}
\caption{Summary of detection channels}
\label{tab:sumchan}
\begin{tabular}{lc}
\hline
$\tau^-$ decay mode & $Br$ (\%) \\
\hline
$e^-\nu_e\nu_\tau$     & 18.0 \\
$\mu^-\nu_\mu\nu_\tau$ & 17.6 \\
$\pi^-\nu_\tau$        & 11.7 \\
$\rho^-(2\pi)\nu_\tau$ & 25.2 \\
$a_1^-(3\pi)\nu_\tau + n\pi^0$  & 14.4 \\
\hline
Total & 86.9 \\
\hline
\end{tabular}
\end{table}

\begin{figure}[ht]
  \epsfxsize=7cm
  \centerline{\epsfig{file=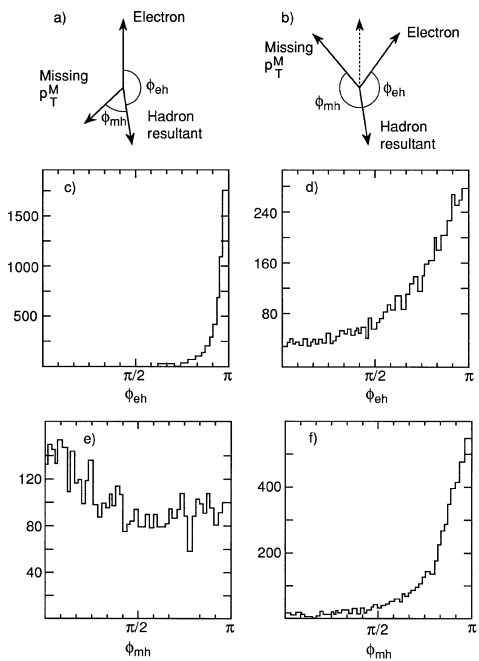}}
  \centering
  \caption {Kinematic variables used in the oscillation
    search for a) $\nu_e$ CC b) $\nu_\tau$ CC; $\tau\rightarrow e$
    in the plane orthogonal to the incident $\nu$. 
    MC distribution of
    \protect$\phi_{eh}$ for c) background d) signal. MC distribution of
    \protect$\phi_{mh}$ for e) background f) signal. See text for
    the definition of angles.}
  \label{phiplots}
\end{figure}

For each event the sum of the momenta transverse to the 
beam direction ,$\Pt$, for all seen particles is calculated
and a resulting missing $\Pt$
vector (\Ptm) is reconstructed.
The angles $\phi_{eh}$ between the electron momentum and
the momentum of hadrons and $\phi_{mh}$
between \Ptm and the hadron vector are defined
in the plane perpendicular to the beam direction
as shown in Fig.~\ref{phiplots}~(a,b). 
In the case of \nue$CC$ background 
simulations show that 
the angle $\phi_{eh}$ is sharply peaked at $\pi$ (Fig.~\ref{phiplots}~(c)).
On the contrary for $\nut$ interactions, the $\tm$ lepton balances the
hadrons, while the electron from \tm decay
acquires a finite $\Pt$ relative to
the $\tm$ direction and will not be back--to--back
with the hadron vector  (Fig.~\ref{phiplots}~(d)). 
So while the \Ptm in the genuine $\nut$ interactions
is due to the missing of the two \nue and \nut \tm decay neutrinos
and therefore is along the \tm direction with a peak near $\pi$
(Fig.~\ref{phiplots}~(f)),
$\Ptm$ in the \nue$CC$ events is expected to 
arise either from missed neutral particles, such as neutrons
and $K^{0}_{L}$, which contribute to an enhancement in $\phi_{mh}$ near
0, or from mismeasurements which will give a flat contribution to 
 $\phi_{mh}$ (Fig.~\ref{phiplots}~(e)). 
It is than expected
that the $\phi_{eh}$ and $\phi_{mh}$ distributions are
quite different for the signal (\nut $CC$) and the background 
(\nue $CC$) events. Similar distributions are used
in the muonic and hadronic decay channels of the $\tm$ to select
the signal.
A two dimensional cut in the plane
($\phi_{eh}$, $\phi_{mh}$) is then defined and
the contamination from all background sources,
$e^-$ background coming from \nue $CC$ and 
$e^-$ background coming from the hadronic jet in \numu $NC$,remaining
after all selection cuts is minimized.
It was evaluated that for an efficiency for the signal
of $13.5 \%$ one expects 4.6 background events
in total.
It is important to stress that most of the backgrounds
can be studied in the data themselves rather than
having to be estimated from Monte Carlo.
The $e^-$ background coming from the hadronic jet in \numu $NC$ can be studied
using \numu $CC$ events and ignoring the $\mu$ .
The background coming from \nue $CC$ can be studied
replacing the $\mu$ in \numu $CC$ events 
with a simulated $e^-$.

\section{Detector performance}
\begin{figure}[hbp]
    \epsfxsize=7cm
  \centering
  \centerline{\epsfig{file=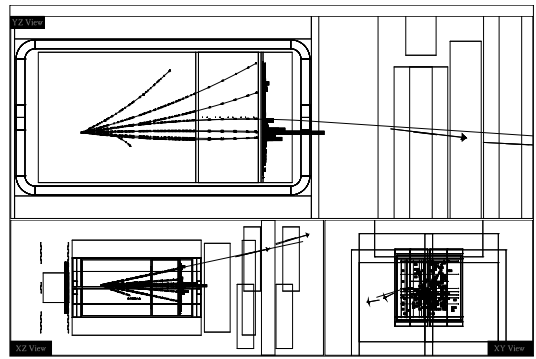}}
  \caption{A reconstructed charged current candidate (1995 run)}
  \label{event_display}
\end{figure}

The NOMAD detector measures and identifies the electrons, muons, photons
and hadrons produced in neutrino interactions. 
A reconstructed charged current candidate (1995 run)
is shown in fig.~\ref{event_display}.
A detailed description of the 
detector is given elsewhere\cite{Rubbiaa}.
The active target consists of
drift chambers (DC) with low
average density ($\approx 0.1$\, g/cm$^{3}$). The measured
single hit resolution is 180 $\mu$m.
These DC are located in a
magnetic field of 0.4~T which allows the determination of the
momenta of charged particles with little degradation due to multiple 
scattering, given the low $Z$ of the chamber material: 
$\sigma_p/p \approx 5\%/\sqrt{L}\oplus 
0.8\%p/\sqrt{L^5}$ ($L$ in m, $p$ in GeV/$c$). 
The active target
is followed by a transition radiation detector (TRD) to tag
electrons, a preshower (PS) detector, an electromagnetic calorimeter 
(ECAL,$ \sigma(E)/E \approx 4\%/\sqrt{E/GeV}$ for electrons)   
a muon absorber and muon chambers.
\par
In the 1994 run all these detectors were functional, but
only two out of eleven drift chamber target modules were installed
behind a provisional nonactive target. More drift chambers
were installed gradually for the 1995 run, and the detector was completed
in August 1995 .
In the beginning of 1995 a hadronic calorimeter\cite{HCAL} 
(HCAL,$ \sigma(E)/E \approx 120\%/\sqrt{E/GeV}$) was installed 
downstream of the ECAL and a front calorimeter 
was implemented upstream of the active target.

\section{Particle identification}
Several detectors contribute to electron identification.
The TRD achieves a pion rejection of 10$^{-3}$ whith an electron
efficiency $ > 90\%$ for $p > 2$ GeV/$c$
as measured in test beams. 
Consistency of the momentum measured 
in the drift 
chambers with the energy and 
shower shape measured by the ECAL/PS
system yields an additional
rejection factor $\geq$ 100.
\par
\begin{figure}[htp]
  \bc
  \hspace{0mm}  \epsfig{file=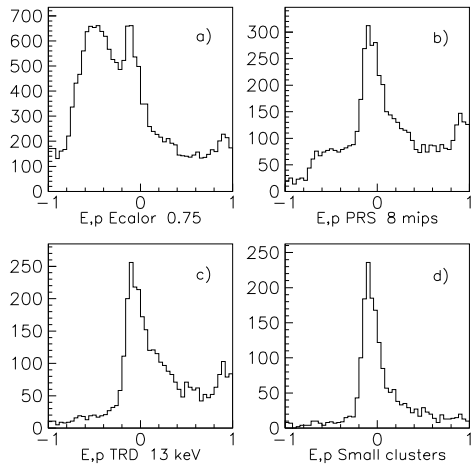,width=70mm}
  \ec
  \caption{The distribution of $(E-p)/(E+p)$ for a) all drift
    chamber tracks which point to a cluster in ECAL with
    energy greater than 0.75 GeV; b) same with cut on
    the preshower energy; c) same with TRD pulse height
    cut; d) same with cut on the size of the ECAL cluster.}
  \label{E-p/E+p}
\end{figure}

Fig.~\ref{E-p/E+p} shows the evolution of this consistency variable after
successive cuts. We interpret the peak near zero as electrons originating 
from photon conversions and \nue CC interactions. 
\par
The low density NOMAD target ( about 1.3 $X_0\,$ in total)
allows the detection of photons and $\pi^0$ in the 
ECAL/PS system
with a good spatial and energy resolution .
Gamma candidates were selected with the following criteria:
i) energy larger than 200 MeV;
ii) no DC tracks
pointing to the cluster (within 15 cm);
iii) a pre-shower signal higher than 4 mip and matched
to the cluster closer than 6 cm ;
iv) less than 4 gamma candidates per event;
v) reconstructed vertex inside the fiducial volume.
\par
\begin{figure}[hbp]
    \epsfxsize=7cm
  \centerline{\epsfig{file=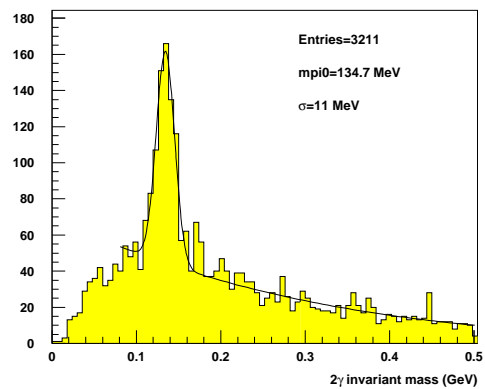}}
  \centering
  \caption{Two photon invariant mass distribution in NOMAD 
            ECAL/PS system.}
  \label{pi0nom}
\end{figure}

A clear $\pi^0$ peak is observed in the measured two photon
invariant mass distribution (Fig.~\ref{pi0nom})
from a sample of 1995 data.
\newpage
\newpage
Charged hadrons appear as tracks in the drift chambers which 
are neither identified as a muon nor as an electron. The measured hadron
momentum and multiplicity distributions 
agree reasonably well with
Monte Carlo predictions.
\par
\begin{figure}[hbp]
  \epsfxsize=70mm
  \centerline{\epsfig{file=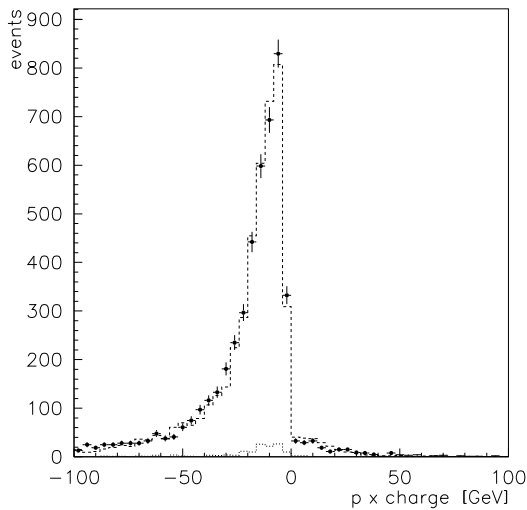}}
  \centering
  \caption{Momentum distribution for both negative and positive muons 
    originating from the inactive target (1994 data). The measured data points 
    are compared to the Monte Carlo prediction (dashed line).}
  \label{muon_mom}
\end{figure}

Muons are identified by drift chamber tracks matching the corresponding tracks 
in the muon chambers. As expected, the measured muon momentum spectrum is
dominated by negative muons from charged current neutrino interactions
(Fig. \ref{muon_mom}). 
The contribution of positive muons is the one expected 
according to the beam contamination.
The distribution in momentum is in agreement
with the MC prediction based on the $\nu$ flux energy distribution.
\section{Conclusion}
The NOMAD detector is working well. The active target
was fully completed in August 1995.
Electrons and muons have been identified and distinguished
from charged hadrons. $\pi^0$'s have also been identified.
NOMAD will continue taking data
until the end of 1997 to search
for \mutau oscillations.

\section*{Acknowledgments}
Thanks are due to the NOMAD technical staffs for
their invaluable support.
\par
I wish to thank the organizers for this very interesting and
fruitful workshop in the nice historic city of Toledo.

\end{document}